\documentclass[12pt]{iopart}
\usepackage{iopams} 
\usepackage{graphicx}
\usepackage{epstopdf}
\usepackage{mathrsfs}
\usepackage{cite}

\begin{document}

\title{Domain dynamics and fluctuations in artificial square ice at finite temperatures.}

\author{Z.~Budrikis$^{1, 2}$, K.~L.~Livesey$^{1, 3}$, J.~P.~Morgan$^{4}$, J.~Akerman$^{4, 5}$, A.~Stein$^{6}$, S.~Langridge$^{7}$, C.~H.~Marrows$^{4}$ and R.~L.~Stamps$^2$}
\ead{zoe.budrikis@gmail.com}
\address{$^1$School of Physics, The University of Western Australia, 35 Stirling Hwy, Crawley 6009, Australia}
\address{$^2$SUPA School of Physics and Astronomy, University of Glasgow, Glasgow G12 8QQ, United Kingdom}
\address{$^3$Department of Physics and Energy Science, University of Colorado, Colorado Springs, Colorado 80918, USA}
\address{$^4$School of Physics and Astronomy, University of Leeds, Leeds LS2 9JT, United Kingdom}
\address{$^5$Instituto de Sistemas Optoelectr\'{o}nicos y Microtecnolog\'{i}a (ISOM), Universidad Polit\'{e}cnica de Madrid}
\address{$^6$Center for Functional Nanomaterials, Brookhaven National Laboratory, Upton, New York 11973, USA}
\address{$^7$ISIS, Rutherford Appleton Laboratory, Chilton OX11 0QX, United Kingdom}

\begin{abstract}
The thermally-driven formation and evolution of vertex domains is studied for square artificial spin ice. 
A self consistent mean field theory is used to show how domains of ground state ordering form spontaneously, and how these evolve in the presence of disorder.
The role of fluctuations is studied, using Monte Carlo simulations and analytical modelling. Domain wall dynamics are  shown to be driven by a biasing of random fluctuations towards processes that shrink closed domains, and fluctuations within domains are shown to generate isolated small excitations, which may stabilise as the effective temperature is lowered.
Domain dynamics and fluctuations are determined by interaction strengths, which are controlled by inter-element spacing. The role of interaction strength is studied via experiments and Monte Carlo simulations.
Our mean field model is applicable to ferroelectric `spin' ice, and we show that features similar to that of magnetic spin ice can be expected, but with different characteristic temperatures and rates.
\end{abstract}

\date{\today}

% \pacs{75.50.Lk, 75.10.Kt, 75.75.Àc, 75.78.Àn}

\submitto{\NJP}

\maketitle

\section{Introduction}
Artificial spin ices \cite{Wang:2006} are constructed as finite arrays of elongated magnetic dots whose magnetisations are assumed to be well approximated by Ising spins. 
The stray magnetostatic fields of each island mediate interactions, which are frustrated by geometry.
These systems are called `ices' because minimisation of the magnetostatic energy leads to behaviour resembling that governed by the ice rule for the ground state of solid water, i.e., two-in, two-out spin configurations \cite{Pauling:1935}.

The geometry of artificial spin ices can be controlled to a large extent \cite{Tanaka:2006, Qi:2008, Mengotti:2008, Li:2010, Li:2010a}, and resulting magnetic configurations can be imaged directly using techniques such as MFM \cite{Wang:2006, Ke:2008, Remhof:2008, Morgan2011, Morgan2011a}, PEEM \cite{Mengotti:2008, Mengotti:2010, Rougemaille2011}, and electron microscopy \cite{Qi:2008, Phatak2011}. 
Most studies have been made at room temperature with relatively large, thermally stable magnetic elements. 
This thermal stability is enforced by choosing island volumes large enough that the barrier to magnetisation reversal, primarily associated with shape anisotropy, is much higher than room temperature,  so that dynamics can only be induced by applying a magnetic field. Spatial studies are then made with the system in a steady state configuration.

Because the focus to date has been on athermal systems, very little is known about how two dimensional spin ice arrays respond to thermal fluctuations. 
Recent reports provide evidence that there are magnetic features visible in as-grown arrays that form during early stages of growth \cite{Morgan2011}, namely long range ground state ordering, with well defined domain walls and small clusters representing configurational excitations above the ground state. 
The origin of the ordering and excitations was argued to be thermal. 
In another recent work, thermal loss of macro-spin order was reported in a square artificial spin ice patterned $\delta$-doped Pd(Fe) film \cite{Kapaklis2011}.

From a theoretical perspective, simulation studies \cite{Libal:2006} of an analogue of square ice constructed from charged colloids in double-well optical traps reveal a freezing transition as the temperature is reduced relative to the height of the barriers between pairs of wells. The same authors also find \cite{Libal2011a} that strong thermal noise is required to alter the hysteretic behaviour of the system. Recently, Monte Carlo simulations of magnetic square ice have been presented \cite{Silva2011}, which indicate a sharp peak in the specific heat and a peak in the density of closed loops of spins flipped against the ground state at approximately the same temperature.

In this paper we discuss configurational dynamics of square artificial spin ices in terms of macro-domain growth and boundary movements at finite temperatures. The picture we present is one in which domain boundaries can flow, and also form channels over which magnetically charged vertices move. We find that a domain of ground state configuration should, in the absence of disorder, spontaneously grow until it fills the array in a finite system.  We show also that thermal fluctuations can accelerate domain boundary movement, and create clusters similar to those observed experimentally.

\begin{figure}
 \centering
   \includegraphics[width=0.7\columnwidth]{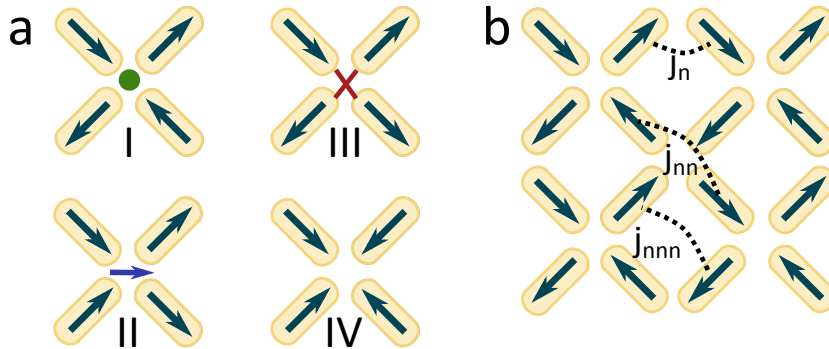}
  \caption{\label{geometry} (a) Examples of each of the four vertex types. The symbols in the centres of vertices I--III are used later in the text to indicate the vertex type.
  (b) Geometry of the square artificial spin ice in this paper. Finite range interactions are assumed, and indicated on the diagram.}
 \end{figure}

We restrict our attention to square array geometries in this work. 
Spin configurations can be completely specified by vertex arrangements, and for the present work, we use  spin or vertex descriptions interchangeably for configurational and energy states. The concept of vertices provides a useful nomenclature for discussing spin configurations \cite{Wang:2006, Moller:2006, Libal:2006, Ke:2008, Libal:2009}, effective temperatures of field-driven demagnetisation \cite{Nisoli:2007, Nisoli:2010} and dynamics \cite{Mol:2009, Budrikis:2010, Mol:2010, Morgan2011a}. We will use vertex descriptions almost exclusively in what follows.
In the square lattice there are sixteen distinct vertex configurations, which can be classified into four groups, Type I---IV on the basis of magnetic charge and total moment. The groups are ordered on the basis of energy. Examples of each type are shown in figure~\ref{geometry}(a). 
The ground state corresponds to a complete chessboard tiling of the two Type I vertices, and is two-fold degenerate. 

%[FIG Geometry & nn,nnn definition]

The organisation of the paper is as follows. In the next section we describe general features of domain dynamics in the presence of quenched disorder, using a mean field theory, in reference to conventional magnetic field and zero field cooling experiments. 
We then discuss effects of thermal fluctuations and strength of coupling using Monte Carlo simulations. These results are compared to experiments.
We also discuss how thermal fluctuations can create small cluster excitations above the ground state, which can be understood using statistical arguments.
Throughout this work, we see that Type I domain formation and growth are general phenomena, a result which is corroborated by simulation results for artificial `spin' ice in ferroelectric media, which we discuss in an appendix.

\section{Magnetisation Processes in Mean Field Approximation}

Magnetic elements are approximated as block Ising spins, and arranged on a finite two dimensional square lattice, as shown in figure~\ref{geometry}(b).  The importance of long range dipolar interactions for correlations is a topic of investigation  \cite{Moller:2006, Rougemaille2011, Moller2009, Chern2011}, but our simulations suggest that shortcomings of the finite range interaction approximations are significant only in idealised perfect systems.
In the models used throughout the main text of this paper, we consider only the three nearest neighbour interactions: $J_{\rm n}$, $J_{\rm nn}$, and $J_{\rm nnn}$ in a point dipole model. 
The square array geometry can be thought of as two sets of parallel lines of elements, with the sets aligned at 90$^{\circ}$ to each other. The $J_{\rm nn}$ accounts for interactions along each line, and the $J_{\rm nnn}$ couples elements from adjacent parallel lines. The $J_{\rm n}$ couples elements from the two different sets. These couplings are indicated schematically by dashed lines in figure \ref{geometry}(b).
In the appendix, we give results for a ferroelectric ice where all point dipoles in the array are summed over, and find qualitatively similar results to those presented in this section.

%- Mean field (MF) model
The energy of a magnetic element is determined by the local interactions and any applied fields, $H$. Denoting the magnetic moment at site $i$ as $m_i$, reversal of $m_i$ will occur when
\begin{equation}
\epsilon_c < \mu_B  m_i  H  +  m_i \cdot \sum_{j\ne i}J_{i,j}\langle  m_j \rangle.
\label{eq:criticalfield}
\end{equation}
Here, $\mu_B$ is the Bohr magneton, $\langle m_i \rangle$ is the mean field thermal average moment at site $i$, and $\epsilon_c$ is an energy barrier to reversal. We suppose that the energy barrier to reversal is proportional to the anisotropy responsible for the Ising like behaviour of the magnetic elements. In the mean field approximation this is
\begin{equation}
\epsilon_c = \frac {1}{2} m_i K \langle  m_i \rangle,
\label{eq:ec}
\end{equation}
where $K$ represents an anisotropy barrier. Each magnetic element is assumed to behave as a single `soft' macrospin, with a total moment that depends upon temperature. We approximate the temperature behaviour of an element's moment by the Langevin function $L(x)$:
\begin{equation}
\langle  m_j \rangle = L \left[ \beta ( \mu_B  H+ \frac {1}{2} K\langle  m_j \rangle + \sum_{k \ne j} J_{j,k} \langle  m_k \rangle ) \right].
\label{eq:langevinavg}
\end{equation}
$\beta$ is the inverse temperature $1/(k_B T)$ where $k_B$ is the Boltzmann constant. 

Previous studies have shown that, in field-driven dynamics, a perfect system allows nucleation of Type I domains on a Type II background through Type III vertex dynamics that start at array edges where moments have fewer neighbours and therefore lower reversal thresholds \cite{Budrikis:2010}. Local variations in the coupling or reversal barrier parameters can facilitate nucleation processes to start inside the array, leading to less sensitivity to the array boundaries \cite{Libal:2009, Budrikis2011networks, Budrikis2011ConstH}. 
In a simple model for switching barriers, the barrier height is proportional to the element volume.  One would expect that during early stages of deposition of material, the element thicknesses are distributed over a range, leading to a spread in switching field values.
Furthermore, significant distributions in switching fields have been observed experimentally in fully-grown, athermal artificial spin ice \cite{Kohli2011, Budrikis2011ConstH, Ladak:2010, Mengotti:2010, Daunheimer2011} and switching field disorder has been shown to give similar outcomes in numerical simulations as other disorder types~\cite{Budrikis2011disorder}. We use this as motivation for describing disorder in terms of the reversal barriers. We assume the barriers vary randomly in an inverval $\Delta$ centred about $K$, that is, $[K - \Delta/2, K + \Delta/2]$.

The mean field model allows one to examine the stability of the ground state relative to temperature but cannot capture correlations such as those involved in avalanche processes, and such processes will be discussed later.  
Initially, all moments are assumed to be of unit magnitude in some specified configuration.
In the present model, an element $i$ is picked at random, and the time averaged reduction of its magnetisation is calculated according to (\ref{eq:langevinavg}). If the magnetisation state becomes unstable, that is, if (\ref{eq:criticalfield}) is satisfied, then $m_i$ is set to $-m_i$.
The process is repeated by choosing a new moment at random, and the iterations continue until a steady state is found such that no $m_i$ value changes by more than $10^{-4}$.

Reduced units used throughout the paper are defined so that $|m_i| \le 1$, $J_n=1.5 J_o$, $J_{nn}=0.7 J_o$, and $J_{nnn}=0.3 J_o$.   The reversal barrier has been chosen to be, in reduced units, $K=10 J_o$. Arrays consist of 640 vertices, and array boundaries are taken to be `open' boundaries for which the edge elements have only three nearest neighbors, and corner elements have two nearest neighbours, as shown in figure~\ref{geometry}(b).

The stability of the Type I ground state can be studied by starting the system at low temperature with a complete Type I tiling, and then increasing the temperature. The critical temperature can be calculated within mean field theory from the average magnetisation of an element in a Type I tiling, which obeys
\begin{equation}
\langle m \rangle = \coth{(\beta h_{\rm loc}\langle m \rangle})-1/{(\beta  h_{\rm loc}\langle m \rangle)}
\label{eq:mfconsist}
\end{equation}
where  $h_{\rm loc}=2(2J_{\rm n}-J_{\rm nn}+J_{\rm nnn})+\frac{1}{2}K$. The critical temperature $T_c$, is given by $1/\beta = h_{\rm loc}/3$, to first order approximation. Using the parameters listed above, $T_c=3.4T_o$, where $T_o = J_o/k_B$.

\begin{figure}
 \centering
  \includegraphics[width=0.5\columnwidth]{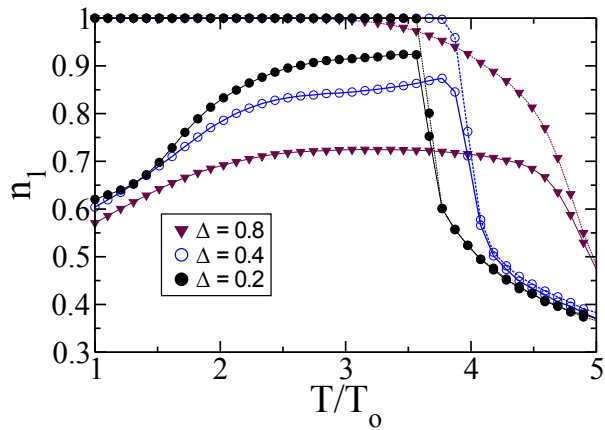}
  \caption{\label{FCZFCT1} Thermal reduction of Type I population starting from a random configuration (solid lines) and starting from a perfect ground state (dashed lines). Disorder is introduced through random variations in individual barriers to reversal. Thermal effects are calculated in a self consistent mean field approximation. The total population of all Type I vertices is shown as a function of reduced temperature for different amounts of disorder. For each disorder value, the evolution is traced starting from low temperature for a uniform single domain of Type I, and a random configuration. The transition temperature approaches the expected value $3.6 T_o$ in the limit of no disorder.  Disorder broadens the transition considerably.}
\label{mfzfcfc}
 \end{figure}

Because changes in magnetic configurations are driven by instabilities only, in the mean field model, the transition is very sharp with a well defined critical temperature in a perfect system without disorder. Disorder broadens the transition region in temperature, and leads to formation of domains of the two different possible Type I tilings separated by Type II and III chains. Examples are shown in figure \ref{FCZFCT1} for Type I populations $n_1$ as functions of temperature for different disorders $\Delta$. The population shown is the sum of both Type I tiling possibilities. Other authors also report a broad melting transition in a thermal ice prepared in a magnetised (Type II vertex tiling) state \cite{Kapaklis2011}.

With disorder, a sharp initial reduction occurs in $n_1$ near the critical temperature, corresponding to the nucleation and growth of domains. This is followed by a long tail in which the total population of Type I vertices approaches the value $n_1=1/8$, corresponding to the high temperature limit in which each of the $16$ possible vertices appear with equal probability. Note that the transition temperature goes to the expected value of $3.4T_o$ as $\Delta \rightarrow 0$.

Also shown in figure \ref{FCZFCT1} is the Type I population that evolves from a random vertex configuration starting at low temperature. This is analogous to the `zero field cooled' state of a ferromagnet that, initially at some temperature above the ordering temperature, has been quenched to low temperature. In the mean field model, a state with large Type I domains immediately evolves at low temperature if $K$ is sufficiently small relative to the $J$'s. The $n_1$ populations grow with increasing temperature as other vertices become unstable.

%[FIG FC/ZFC type 1]

\begin{figure}
 \centering
 \includegraphics[width=0.9\linewidth]{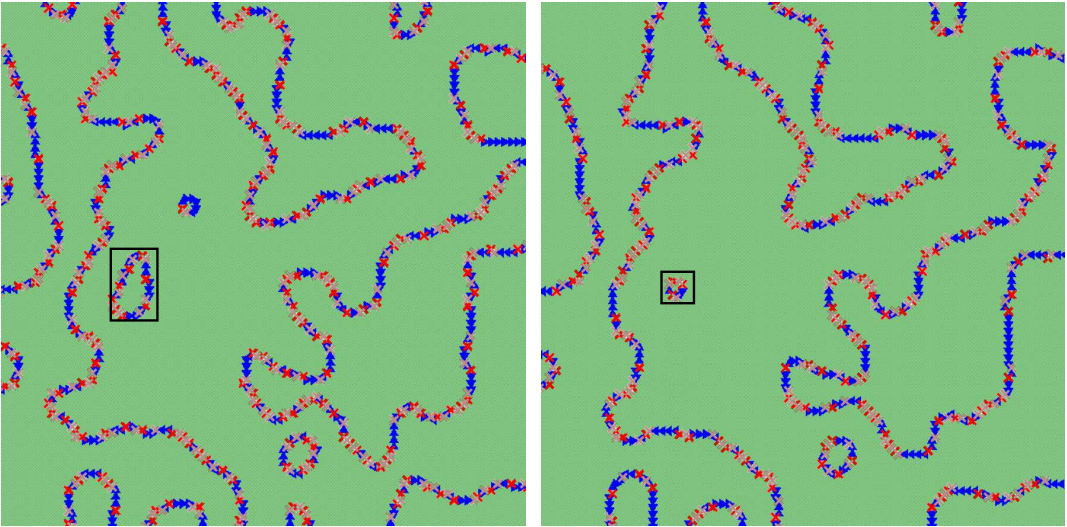}
   \caption{\label{meanf1} Example configurations during numerical iteration of the self consistent mean field algorithm. Green regions are Type I domains. Blue arrows indicate Type II vertices, and point in the direction of the vertex net moment. Red crosses are Type III vertices. The left panel shows a configuration found after $100$ iteration loops begun from a random configuration at temperature $T_o$ and no disorder. The right panel is generated $100$ iteration loops later. The box in each panel indicates a closed domain which has shrunk over the course of the evolution.}
 \end{figure}

This growth of domains occurs through the elimination of small domains and clusters. An example is shown in figure \ref{meanf1}, where vertex configurations at two steps during the numerical iteration are shown. The temperature is $T_o$, and disorder $\Delta=0.1$. Green squares represent Type I vertices. The blue arrows are Type II vertices, and their direction indicate the orientation of the local Type II moment. The light and dark red crosses represent the two charge flavors of the Type III vertices. 

The first image (left) is taken after $100$ iterations after an initially random configuration of moments. Already a well defined Type I domain structure has formed. 
The domains are separated by chains composed of Type II and III vertices with net magnetic moment and charge, and the energy of their formation acts as a surface tension on the Type I domains. For this reason, the chains tend towards straight in the limit of no disorder, since corners lengthen the chain and can involve Type III's. 
The second image (right) shows the evolution after $100$ additional iterations. The longest chains have shifted slightly, and positions of the Type III vertices within the chains have changed. Most notably, the smallest domains have vanished, and other small domains are reduced in size. Note that Type IV vertices are highly unstable, and do not survive past the first few iterations.

%[FIG stages of domain growth after FC (type 1) for different tempertures]

\begin{figure}
\centering
  \includegraphics[width=0.7\columnwidth]{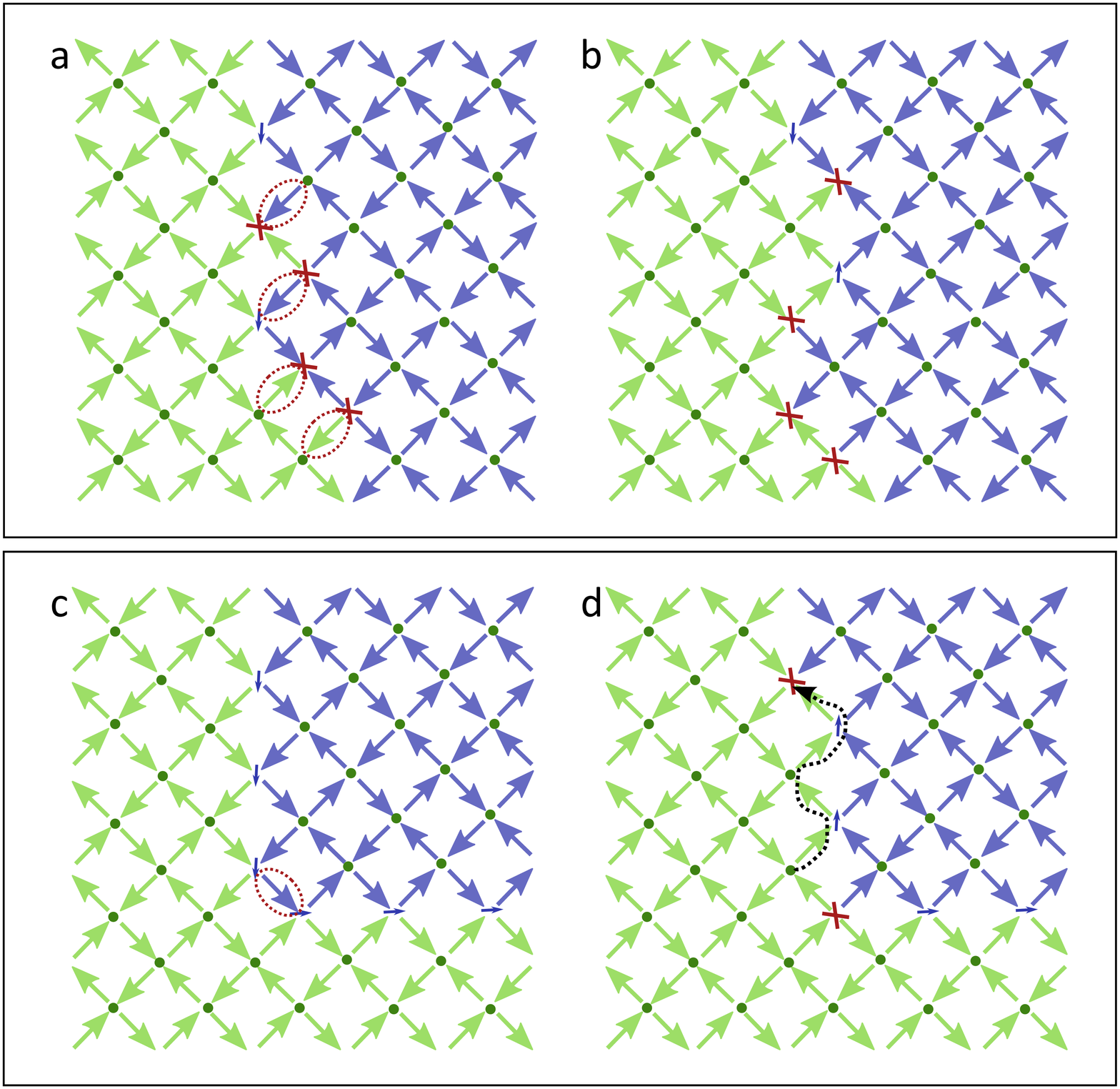}
  \caption{\label{dw_motion} (a, b) Random domain wall fluctuations, driven by Type III vertex motion. The circled spins in (a) can flip at approximately zero cost, to yield the configuration shown in (b). (c, d) Domain shrinking by Type III vertex creation and propagation. The corner of the domain wall provides a Type III nucleation site, shown as the circled spin in (c). Once nucleated, the Type III vertex can propagate at approximately zero energy cost, shrinking the darker blue domain, as seen in (d).  The arrows representing spins are colour-coded light green and dark blue according to which Type I domain they belong to. Vertex types are indicated by symbols: Type I, II and III vertices represented as circles, arrows and crosses, respectively.}
\end{figure}

The reduction in size and eventual annihilation of small domains is the result of an energetic biasing of random domain wall motion towards motion that makes walls shorter by moving closed walls inwards.
Dynamics on a Type I/Type II vertex background -- such as domain wall motion -- can be described in terms of Type III vertex motion \cite{Mol:2009, Budrikis:2010, Morgan2011a}. There are therefore two energy costs associated with dynamics: the cost of creating Type III vertices, and the cost of their propagation. This latter cost is approximately zero in processes where the vertex populations are conserved.  For example, in figure~\ref{dw_motion}(a), Type III vertices on the domain wall can move at approximately zero cost by flipping the circled spins, to yield the configuration shown in figure~\ref{dw_motion}(b). Accordingly, the lowest-energy mode of domain wall motion is the propagation of Type III vertices that have formed at random positions during wall creation. These random fluctuations  do not favour either growth or shrinking of domains on average. The existence of random fluctuations lead us to speculate that an analogy to creep motion~\cite{Lemerle1998} may be possible.

The cost of Type III vertex creation depends on the local spin configuration, and is lowest at corners in the domain walls. For example, in figure~\ref{dw_motion}(c), the easiest spin to flip is the circled spin at the domain wall corner, which can flip to create a Type III vertex pair. 
The resulting Type III vertices can propagate at approximately zero cost by moving along the domain wall, flipping a diagonal chain of spins on the inside of the wall. This process moves the domain wall inwards, as illustrated in figure~\ref{dw_motion}(d). Such sequences of Type III creation and propagation are the domain wall motion mechanism with second-lowest energy cost. Thus, over time, the wall motion is biased so that despite random fluctuations, all closed domains will disappear.

%domain wall motion mechanism

\begin{figure}
 \centering
 \includegraphics[width=0.5\columnwidth]{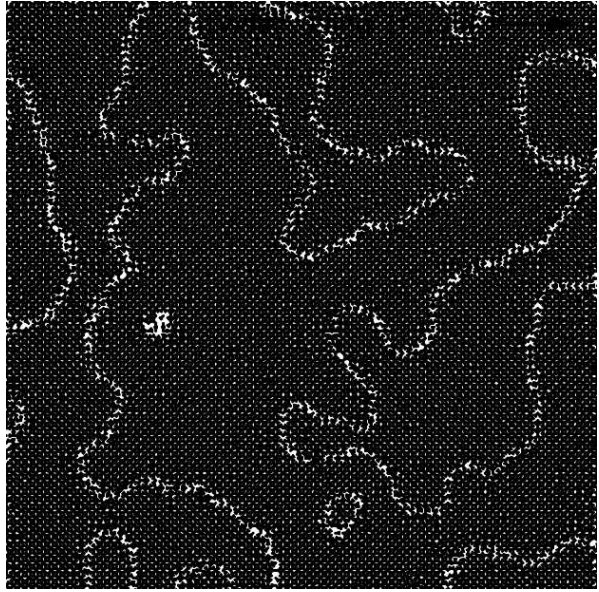}
  \caption{\label{meanfv} The reduction of magnetic moment at finite temperatures is largest within Type II and III chains bounding Type I domains. Here a greyscale plot of the moment is shown for the second example configuration of figure \ref{meanf1}. Dark signifies large moment, and white signifies small moment. The dark regions correspond to Type I domains, whereas the moment is dramatically reduced in the domain walls.}
 \end{figure}

As a final comment on the mean field results, the local fields along the element magnetisations are weakened within the chains, compared to the fields in regions of lower-energy Type I vertices. In the mean field model this leads to a suppression of the magnetisation in the elements participating in the chains.
The magnitude of the suppression depends upon the temperature, and can be very large as one nears $T_c$.
This effect is shown by the vector magnitude map in figure \ref{meanfv}. The magnitude of element magnetisation is shown in grayscale, with black being unity and white being small. The domain pattern corresponds to the configuration shown in the second panel of figure \ref{meanf1}.  One sees that the moment is considerably reduced within the chains, and especially for small clusters. We note that instabilities occur when the moment of individual elements vanish, and these are most likely to occur in the Type II walls.

% reduced moments in chains

\section{Fluctuations and Disorder}

The energy differences between configurations are often small, even for configurations that are very different. For this reason, single spin flips can drive avalanches that lead to large configurational changes and strongly modify vertex populations as domains evolve.  To capture these dynamics, we turn to Monte Carlo simulations.

The model is based as before on the point dipole approximation assumed for the mean field model. The form of the energy used is the same, but evolution is  described using a heat bath algorithm, rather than a simple critical field defined by (\ref{eq:langevinavg}).  Additionally, we consider the case of a large Curie temperature so that the effective block moments are insensitive to temperature. In this limit the energy used for the heat bath algorithm is
\begin{equation}
\epsilon =  m_j  ( \mu_B H + \frac{1}{2} K m_j+ \sum_{k \ne j} J_{j,k} m_k ),
\label{eq:mcenergy}
\end{equation}
where now all magnetisations have fixed magnitude as in an Ising model, with $|m_i|=1$. The barrier to reversal is determined by the anisotropy constant $K$. The parameters used are the same as those for the mean field model. The results shown below were made with $10,000$ Monte Carlo steps at each temperature and averaged over $20$ realisations of disorder at each disorder strength.

\begin{figure}
 \centering
\includegraphics[width=0.5\columnwidth]{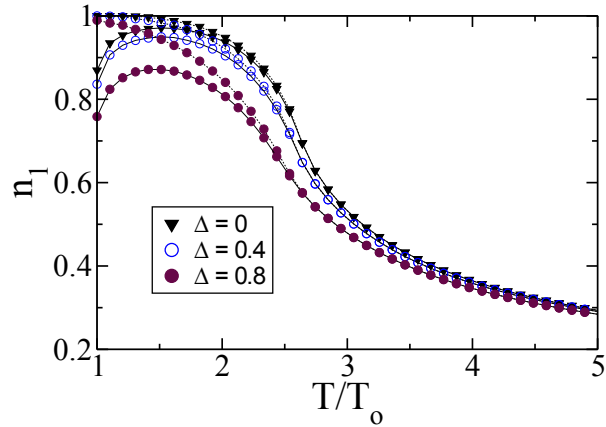}
  \caption{\label{mczfcfc} Thermal fluctuation effects in an artificial square ice calculated using Monte Carlo, and shown as a function of temperature. The total population of all Type I vertices is shown as a function of reduced temperature for different amounts of disorder. For each disorder value, the evolution is traced starting from low temperature for a uniform single domain of Type I (dashed lines), and a random configuration (solid lines). At high temperatures the population tends towards $1/8$, the value expected for a random sampling of the $16$ possible vertex types.}
 \end{figure}

As in the mean field calculation, the role of disorder is largely to broaden the transition. Example heating curves are shown in figure \ref{mczfcfc}, where thermal evolution of ground state and random tilings are shown, for different disorder strengths $\Delta$. The most striking feature, in comparison to the analogous mean field results shown in figure \ref{mfzfcfc}, is the smooth behaviour of the population at the transition, and the rounding in the transition region.

The dependence on disorder is also less dramatic than in the mean field case. In both mean field and Monte Carlo simulations, the spatial distribution of switching barriers is weakly connected to the final distribution of domains and chains. An analysis of the distribution of $K$ values shows that there is a propensity for chains to locate on strong pinning sites, where $K$ is large, and for weak pinning sites to lie more frequently inside domains. This result can be expected because domains grow via thermally activated spin flips. Growth is accomplished through distortions of chains, and chains will pin at sites where $K$ values are too large to allow thermally driven element magnetisation flips. Also, as a consequence, the final resulting domain configuration at a given temperature displays rough chains for large $\Delta$ values, and flat segmented chains for low $\Delta$ values.

%%FIG FC/ZFC MC results
 
Differences between Monte Carlo and mean field results appear because of how thermal fluctuations drive domain growth. 
Fluctuations allow the system to explore dynamical pathways different to those dictated by a stability analysis, and are therefore quite significant in determining the final domain configuration. 
At temperatures below the transition, the dominant paths tend towards the lower energy, single domain state.

\begin{figure}
 \centering
 \includegraphics[width=0.9\linewidth]{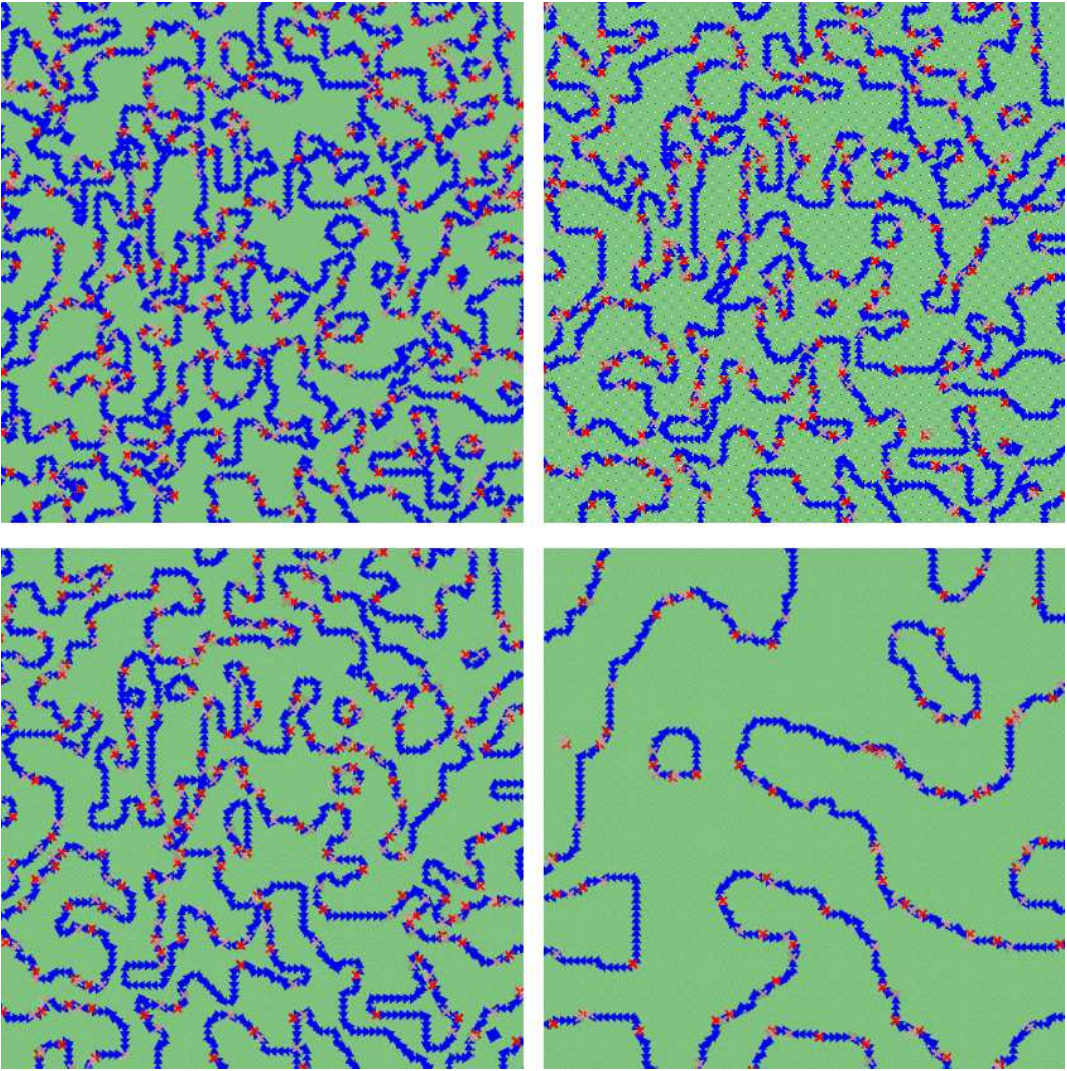}
  \caption{\label{mcevolve} Different stages in evolution of a vertex configuration as calculated using Monte Carlo simulations. Green regions are Type I domains. Blue arrows indicate Type II vertices, and red crosses are Type III vertices. The first three panels show configurations after $1,000$, $2,000$, and $3,000$ steps  for a temperature of $T_o$ and disorder $\Delta=0.1$. The initial state was random. The final panel shows the configuration after $10,000$ steps. Note that small clusters and domains disappear as large domains grow at the expense of smaller domains. }
 \end{figure}

An example evolution is shown in figure \ref{mcevolve}, using the colour scheme defined in figure~\ref{meanf1}. Four configurations are shown at constant temperature $T=T_o$, starting with a random initial configuration. The first three configurations are taken after $1000$, $2000$, and $3000$ steps. The fourth configuration (in the bottom right hand corner) is taken after $10,000$ Monte Carlo steps. The amount of disorder in this example was small relative to both $K=10J_o$ and $J_{\rm n}=1.5J_o$, with $\Delta=0.1$. One sees the slow growth of one Type I domain at the expense of another, through the motion and deformation of Type II and III chains. The dynamics is equivalent to that seen with the mean field model. Note that here also Type IV vertices do not survive the first few Monte Carlo steps unless interactions are weak (as with large spacings).

%%FIG example MC evolution
 
   \begin{figure}
 \centering
 \includegraphics[width=0.5\columnwidth]{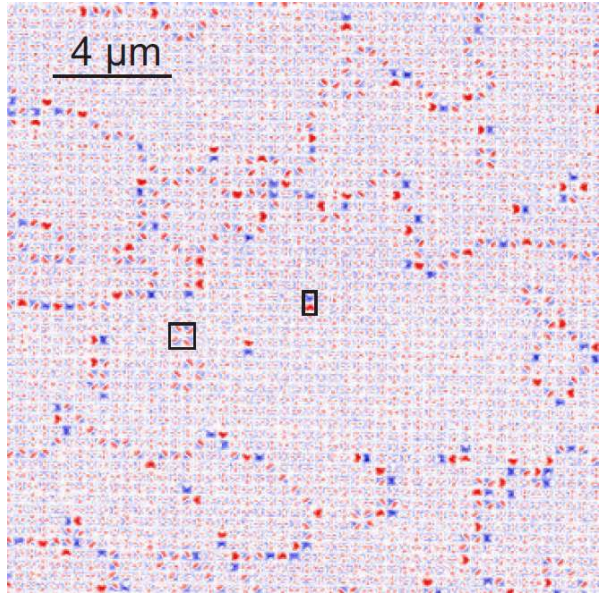} %exp_config.eps}
  \caption{\label{exp_config} MFM image of an as-grown square artificial spin ice, with lattice constant 400 nm, relative to island dimensions of 280 nm $\times$ 85 nm. Red and blue colouring indicate magnetic poles, and domain walls and small excitations are clearly visible on a ground state background. Two examples of small cluster excitations are indicated by boxes. The image has been false-coloured using the software package WSxM~\cite{Horcas2007}. Note that islands in this image are at 45$^{\circ}$ to the islands in other figures in this paper.}
 \end{figure}

In addition to our mean-field and Monte Carlo simulations, we have imaged vertex configurations in as-grown samples, and found domain structures similar to those in simulations. An example MFM image is shown in figure~\ref{exp_config}, where domain walls, which carry net moment are clearly visible on the approximately neutral background of Type I domains. The sample imaged here was previously studied in terms of ground state ordering and thermal excitations~\cite{Morgan2011}, and consists of an array of nominally $280\times85\times26$ nm$^3$ Permalloy islands, with a 3 nm Ti buffer and a $2.5$ nm Al cap.
Full details of sample fabrication are given elsewhere~\cite{Morgan2011}, but the key point is that islands are formed by deposition of Permalloy through gaps in a nanopatterned resist mask, so that island thicknesses -- and hence barriers to magnetisation reversal -- grow over time.

For the simulation, the values shown in the figure were obtained after a finite number of Monte Carlo steps, and the effect of temperature, relative to the local coercive field, is to control the rate at which $n_1$ changes. In the experiment, the deposition rate relative to temperature probably controls the rate at which $n_1$ changes.  We suggest that domain configurations found in the simulation and the experiment can both bo interpreted as snapshots in time of a thermally driven evolution. We note that a difference between the simulations shown in this example and the experiment does exist, in that the as-grown samples support small clusters of flipped spins. We discuss  this further in section~\ref{fluctuations}.

%dependence on spacing: interaction strength
\subsection{Spacing and interaction strength}
\label{spacing}
The preponderance of Type I vertices in domain structures is a consequence of the higher energies needed to form other vertex types, and the corresponding lower probability for their creation. The probability for reversing a moment depends upon the size of local fields relative to thermal energy. This ratio also determines the rate at which domains can grow. 

\begin{figure}
 \centering
 \includegraphics[width=0.8\columnwidth]{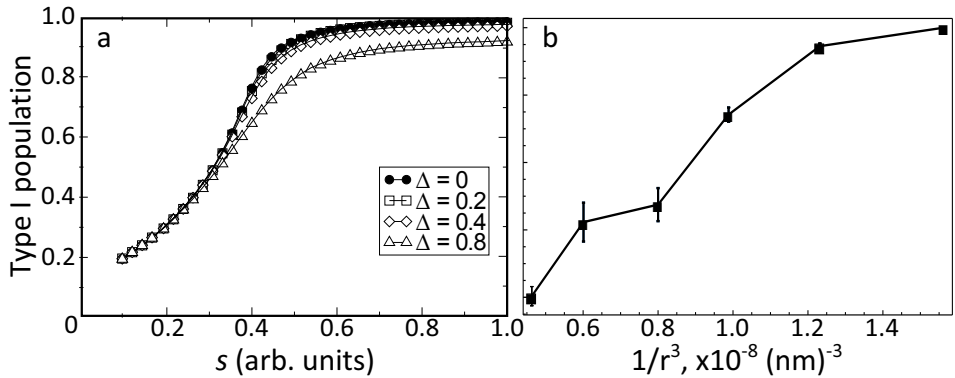}
  \caption{\label{N1vsS} (a)~The  average number of Type I vertices as a function of interaction strength scaling $s$ at a temperature of $T/T_o =1$ for different strengths of disorder as simulated using Monte Carlo. Increasing disorder leads to greater stability for other vertex types relative to the ground state. The $n_1$ population tends toward unity as the simulation time is increased in all cases, but very slowly when disorder is strong. (b)~Experimental results for Type I populations as determined from as-grown samples for arrays grown with different lattice spacings. The populations are shown as a function of $1/r^3$, where $r$ is the lattice spacing, which is proportional to the interaction strength. The populations are the averaged over a number of images (typically 5 or 6) taken across each continuous pattern of 0.5 mm by 0.5 mm total area, and the errors are the standard errors over these.}
 \end{figure}

An example from Monte Carlo simulations is shown in figure \ref{N1vsS}(a). Here the Type I population is shown as function of interaction strength. The interaction strength is scaled by a factor $s$ such that $J_\alpha \rightarrow s J_\alpha /J_o$ for each of the three interactions $\alpha$. Temperature is fixed at $T_o$ in all cases. Several different disorder strengths were studied and the simulations were run for $100,000$ steps at each spacing. The $n_1$ population tends to unity as the interaction strength increases, although the rate at which this occurs depends on disorder strength. For strong disorder, the rate is slowest. We note also that there were no significant Type IV populations observed, although the population did increase with increased disorder.
 
We have also measured statistics of the vertex type populations for as-grown samples with different lattice constants. In these studies, the dimensions of individual elements were held constant at $270\times115\times25$~nm$^3$, but the lattice constant was varied from 400~nm to 600~nm. In this way the interaction strength should decrease, roughly like $1/r^3$ where $r$ is the lattice constant. Results are shown in figure \ref{N1vsS}(b), where the $n_1$ population is given as a function of $1/r^3$.
All samples in the series consist of $0.5\times0.5$~mm$^2$ continuous arrays of Permalloy islands  grown on a 2~nm Ti buffer, with no capping layer, and were fabricated in a single batch, to ensure that fabrication parameters were consistent across the series and so avoid problems with quantitative reproducibility (such as those mentioned in \cite{Morgan2011}).

Both figures \ref{N1vsS}(a) and (b) show that the Type I populations increase with increasing interactions. 
In principle one should be able to extract estimates of interaction energies as a function of spacing from the experimentally measured $n_1$.  However, inspection of figure \ref{N1vsS}(a) shows that one also needs to know the amount of disorder before a prediction for $n_1$ can be made.  The disorder can be expected to be strongly dependent on details of the sample growth and design \cite{Daunheimer2011}.
 
\subsection{Fluctuations and clusters}
\label{fluctuations}
\begin{figure}
\centering
  \includegraphics[width=0.4\columnwidth]{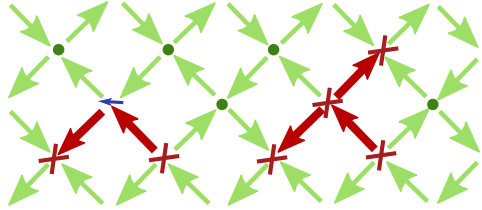}
  \caption{\label{small_excitations} Example 2 and 3 spin-flip cluster excitations on a Type I background. The arrows represent island moments, with the flipped spins shown as bold red arrows. The vertex configurations are also shown with Type I, II and III vertices represented as circles, arrows and crosses, respectively.}
\end{figure}

The role of disorder is subtle. On the one hand, disorder in $K$ leads to randomness in the chains. On the other, disorder also facilitates fluctuations and leads to appearance of small clusters of Type II and III excitations, examples of which are illustrated in figure~\ref{small_excitations}. 
In Monte Carlo simulations, these fluctuations usually appear and disappear quite rapidly (in Monte Carlo step time), but can also lead to structures that may persist through several Monte Carlo steps. 
Experimentally, as-grown samples are seen to exhibit frozen-in fluctuations, as seen in figure~\ref{exp_config}. 
In previous studies, it was shown that the population of clusters decays approximately exponentially with their energy,  which was given as evidence of thermally-driven processes occurring during sample growth \cite{Morgan2011}. In simulations of field-annealed square ices realised in nanostructured superconductors, small structures can only exist independently of domain walls when sufficient disorder is present \cite{Libal:2009}.

An analytical description of the nucleation, growth and decay of an isolated cluster gives insight into the distribution of experimentally observed clusters. As seen in figure~\ref{small_excitations}, clusters can be thought of as connected lines of spin flips against the ground state. The lines may branch. 
In our analytical model, the growth and decay of clusters proceeds by single spin flips that extend or shorten these lines. For example, the left hand cluster shown in figure~\ref{small_excitations} can evolve in to the right hand cluster, and vice-versa. For simplicity, we neglect disjoint clusters: large clusters cannot evolve into two smaller clusters. 

Under the assumption that clusters are excitations above the ground state that are nucleated by random thermal spin flips, we start with an initial perfect ground state and study the temperature-dependent evolution of small clusters of up to four spin flips. The clusters may be grouped by their topology (and hence energy). There are twenty five groups of such clusters: apart from the ground state, there is one distinct cluster of one flip, two clusters of two flips, five of three flips and sixteen of four flips. These groups include clusters that contain Type IV vertices, but these high-energy clusters are suppressed relative to those containing only Types II and III vertices.

The processes of cluster growth and decay are transitions between the groups of clusters, and can be described by a master equation for the probability, $P(A)$, of a cluster having topology $A$:
\begin{eqnarray}
\label{master_equation}
d P(A, t)/dt = \sum_B \biggl(& G(B \to A) \nu(B \to A) P(B, t) \nonumber \\
& - G(A \to B) \nu(A \to B) P(A, t) \biggr),
\end{eqnarray}
where the sum runs over all cluster topologies that a cluster of type $A$ can evolve to or from by a single spin flip. $G(A \to B)$ is a multiplicity that takes into account the number of `pathways' from $A$ to $B$, and depends on the numbers of orientationally-distinct clusters with a given topology and the number of clusters of topology $A$ that can evolve into a particular cluster of topology $B$. The rate $\nu(A \to B)$ is given by the Arrhenius-law factor
\begin{equation}
\nu(A \to B) = f \exp(-(\epsilon_c - \epsilon_{\rm dip})/k_B T).
\end{equation}
$\epsilon_c - \epsilon_{\rm{dip}}$ gives the energy cost of a spin flip to grow or shrink a cluster on a Type I background. $\epsilon_c$ is an arbitrary anisotropy barrier whose exact value does not affect the steady state cluster probabilities. In this model, disorder in anisotropy barriers is neglected, and $\epsilon_c$ is constant always. $\epsilon_{\rm{dip}}$ is the energy of the spin prior to flipping, due to point-dipole interactions among near neighbours ($J_{\rm n}$, $J_{\rm nn}$ and $J_{\rm nnn}$). The attempt frequency $f$ does not affect the steady state probabilities. 

 \begin{figure}
 \centering
 \includegraphics[width=0.5\columnwidth]{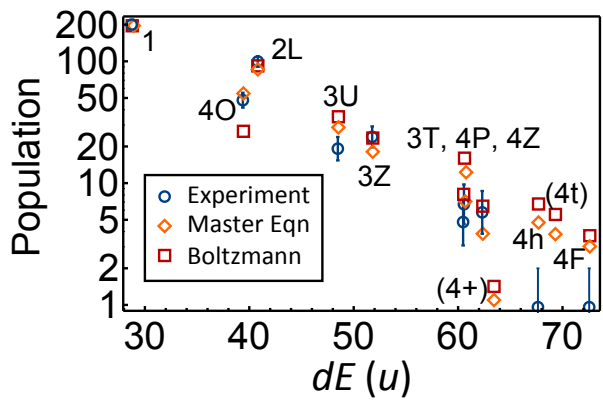}
 \caption{\label{pop_vs_E}  Population \textit{vs} energy above the ground state (in units of $u=\mu_0 m^2/(4\pi r^3)$, where $m$ is the island moment and $r$ is the lattice spacing) for cluster excitations of up to 4 spin flips. As indicated in the legend, data are shown from experiment (these were previously published in \cite{Morgan2011}), the master equations (\ref{master_equation}) and Boltzmann factors. The labels indicate the mnemonics used in \cite{Morgan2011}. The labels 4$+$ and 4t are in parentheses to indicate that those clusters were not observed experimentally in \cite{Morgan2011}. Error bars on the experimental data correspond to the square root of the population of each cluster type.}
 \end{figure}

The twenty five coupled equations of the form (\ref{master_equation}) can be solved numerically, with an initial condition that the system is in the ground state: $P(\varnothing, t=0) = 1$, $P(i, t=0)=0, \forall i \ne \varnothing$, where $\varnothing$ represents the ground state.
 % the ground state has probability $1$ and all other clusters have zero probablity. 
The steady state probabilities of all clusters that do not contain Type IV vertices, normalised to the experimentally observed population of the 1 excitation (that is, a single spin flip), are plotted in figure~\ref{pop_vs_E}, as a function of their energy difference with the ground state (as given in \cite{Morgan2011}).  The agreement between theory and experiment is remarkable.

The figure shows the solution at $k_B T=2.96 u$,  the temperature at which the exponential decay that best fits the theoretical results matches the exponential decay of experimental data. The energy scale $u=\mu_0 m^2/(4\pi r^3)$, where $r$ is the lattice spacing and the moment $m = M V$, where $M=860\times10^3$ Am$^{-1}$ is the magnetisation and $V$ is the island volume.
The clusters that contain Type IV vertices are not shown, but apart from two exceptions they all have populations less than half of the lowest population shown. This is consistent with experiments~\cite{Morgan2011}, where such clusters are not seen and it was concluded such clusters were highly improbable. The exceptions are a Type III -- Type IV -- Type III line and a Type III -- Type IV -- Type II -- Type III cluster, which both have predicted populations of $\sim2$.

For comparison, the figure also shows populations predicted from Boltzmann factors of the form $g \exp(-dE/k_B T)$. 
The degeneracy $g$ of a cluster topology is determined by the number of distinguishable ways  a particular shape can be rotated and reflected, as well as a factor of 2 (equal for all clusters) to account for the possibility of a global spin flip. $dE$ is the cluster's energy above the ground state, taken from \cite{Morgan2011}. The ratio of interaction strength to temperature has been tuned to match the exponential decay to that of the experimental data, giving a ratio of $k_B T = 8.18 u$.

In experiments, configurations are frozen in as the island volume becomes large enough that shape anisotropy barriers suppress island magnetisation reversal. The ratio of temperature to nearest-neighbour coupling can be compared for the master equations and the Boltzmann distribution by estimating the thickness at which freezing-in occurred from the two models. If we estimate the temperature during growth to be 350 K, then the nearest-neighbour interaction is $2.4\times10^{-21}$ J for the master equation and $5.9\times10^{-22}$ J for the Boltzmann factors. For Morgan \textit{et al.}'s $280\times85$ nm$^2$ islands, the estimated interaction strengths correspond to thicknesses of $3.5$ nm (master equation) and $1.7$ nm (Boltzmann factors). Both these estimates are of the same order of magnitude as the estimate of $\sim1$ nm given in \cite{Morgan2011}.

Interestingly, while the Boltzmann factors and the solutions to (\ref{master_equation}) both agree well with the experimental data, the Boltzmann factors agree best for the excitations 1, 2L, 3Z, 4Z while the solutions to (\ref{master_equation}) fit better for the other clusters. 
For example, the `pathway' $1 \to {\rm 2L} \to {\rm 3U} \to {\rm 4O}$ leads to a stable configuration, 4O. In (\ref{master_equation}), we have truncated the maximum cluster size to 4 flips, making 4O very stable. However, even without this approximation, the energy cost of adding an extra spin flip to 4O is high, and the cluster is stable. Thus, the 4O cluster is in some sense a `sink' of probability, and has higher probability than would be expected from Boltzmann factors, which is captured by the master equations.

Finally, we note that the master equations (\ref{master_equation}) explicitly assume an initial ground state configuration, and treats small excitations as having nucleated and grown on this background before being frozen by increasing island switching barriers. The clusters in the as-grown samples of Morgan \textit{et al.} might also be interpreted as regions which have fallen out of equilibrium during the effective cooling of an initially random configuration. 
While we have not explicitly tested such a possibility, the agreement of the master equations with experiment gives further evidence for the picture of nucleating and growing clusters. 

\section{Conclusion}

Vertices in artificial spin ice are local configurations of magnetisations that can be thought of to some extent as classically emergent objects analogous to quasiparticles, able to move and interact according to well defined rules dictated by the lattice geometry. We have shown that domains of ground state vertices form spontaneously in the square lattice. Domain boundaries are defined by chains of Type II vertices, along which Type III vertices can appear, move and drive growth of one domain at the expense of another. 

We have shown that these domain dynamics occur with rates that depend on temperature and the strengths of interactions between elements. Effects of disorder have been studied, and shown to affect mainly the average size and growth rate of domains, while not modifying significantly the fundamental processes involved. A comparison to experimental results was obtained from as-grown samples for which interelement spacing was varied. The experiment and simulations are essentially different in that element thicknesses are changing with time in the experiment whereas temperature is changed in the simulation. Nevertheless, similarities between populations for the Type I ground states were observed between experiment and simulation, suggesting that the as-grown samples are behaving somewhat as frozen snapshots of magnetic ordering within the arrays occurring as described by the simulations presented here.

In addition to domain wall motion, there is also the possibility that an element will flip within a domain. Reversal of a single element in a Type I configuration creates a pair of oppositely charged Type III vertices. Additional reversals lead to clusters of Type II and Type III vertices, that may persist for some time.
We have calculated the probabilities of occurrence of small clusters as a function of their energy, and shown that one obtains a distribution in quantitative agreement with that reported recently from experiments~\cite{Morgan2011}.

Lastly, we have shown that all features of domain formation and growth can be obtained with a generic mean field model. This leads us to suggest the possibility of creating artificial spin ice using ferroelectric media, and we have provided an example in the appendix using parameters appropriate to bulk PbTiO$_3$. In this example we also showed that models using full dipole sums over a lattice of point dipoles produces the same qualitative results as a model using a severely truncated sum.

\ack{The authors thank the Australian Research Council and University of Glasgow for support. Z.B. acknowledges funding from the Hackett Foundation. J.P.M. and C.H.M acknowledge EPSRC and the Centre for Materials Physics and Chemistry at STFC for funding. Research carried out in part at the Center for Functional Nanomaterials, Brookhaven National Laboratory, which is supported by the U.S. Department of Energy, Office of Basic Energy Sciences, under Contract No. DE-AC02-98CH10886.}

\section*{References}
% \bibliography{thermal_ice_refs}
\providecommand{\newblock}{}

\section*{Appendix: Ferroelectric Ice}

 Our discussion so far has been within the context of magnetic spin ice, although our results have general applicability to a number of different systems. We illustrate this with a return to the mean field model, and discuss now domain growth within the context of Landau-Ginzburg theory for ferroelectrics.

 A second order Landau-Ginzburg model is used to calculate the electric dipole moment of an island $i$, given the effective electric field $\bf{E}$ at that point. Each dipole is assumed to lie along only one axis, parallel to the long axis of that island. The ferroelectric free energy density $F$ at island position $\bf{r}_i$ is given by:
 \begin{equation}
 F(i)= \frac{\alpha}{2}P^2(i)+\frac{\beta}{4}P^4(i)-{\bf E}(i) \cdot {\bf P}(i),
 \label{eq:lg}
 \end{equation}
 where $\bf{P}$ is the electric polarisation, $\alpha=A(T-T_c)$ and $\beta$ are the Landau parameters for the material. ${\bf P}(i)$ is assumed to lie parallel with the element long axis. A point dipole approximation is again used with the electric dipole associated with island $i$ denoted as ${\bf p}_i = V {\bf P}(i)$, where $V$ is the volume of an element and ${\bf P}(i)$ is assumed to be uniform across the element.

 Unlike the previous mean field theory, in this model we calculate the electric field at position $i$ by superposing contributions from all other elements (approximated as point dipoles) in the square lattice. The number of elements used is sufficiently small that one can use the simple summation
 \begin{equation}
 {\bf E}(i') = \sum_{i \neq i'} \frac{1}{4 \pi \epsilon_0 \epsilon} \left( 3\frac{(\mathbf r_{i'}-\mathbf r_i)[(\mathbf r_{i'}-\mathbf r_i)\cdot\mathbf p_i]}{|\mathbf r_{i'} - \mathbf r_i|^5}\\
  - \frac{\mathbf p_i}{|\mathbf r_{i'} - \mathbf r_i|^3} \right).
 \label{eq:sum}
 \end{equation}

 Minimisation of (\ref{eq:lg}) will, for a range of ${\bf E}$ values, give two solutions for $p_i$ with opposite sign. One can show simply from (\ref{eq:lg}) that the energy barrier separating the two solutions disappears when
 \begin{equation}
 \frac{\mathbf E(i)\cdot \mathbf p_i}{p_i} > \frac{2\alpha^{3/2}}{\sqrt{27}\beta^{1/2}}.
 \label{eq:ecrit}
 \end{equation}
 This is the condition for stability of an element's polarisation orientation. When this condition is fulfilled, the polarisation is reversed. As in the previous mean field model, this procedure is iterated through a system of elements at a fixed temperature until a steady state is reached.

 Parameters used for the simulations are: $T_c = 1100$~K, $A =7.5 \times 10^5$~C$^{-2}$~m$^2$~N~K$^{-2}$, and $\beta = 2.4 \times 10^9$~C$^4$~m$^6$~N. These parameters give a polarisation in zero field and at room temperature $P=\sqrt (\alpha/\beta)=0.5$~C/m$^2$, which is typical for PbTiO$_3$ in bulk \cite{Haun} and is quite accurate for elements that are over 100~nm long. It should be noted that the Landau expansion (\ref{eq:lg}) should strictly speaking be to sixth-order to most accurately model the ferroelectric phase transition \cite{Haun} but we neglect the $P^6$ terms here for simplicity. 
 Results are given for a $20$ micron square sample with $N=50$ islands along one
 edge ($4900$ islands and $2401$ vertices).
 A spacing of $400$ nm between island centres is assumed. The
 starting configuration for an iteration is taken to be islands having polarisation with constant magnitude $P_i =(\frac{\alpha}{\beta })^{1/2}$, with random alignments along element axes.

 The energy barrier with these parameters for
 flipping an element polarisation is much larger than the
 thermal energy, for most temperatures below the Curie point, so reversal via thermal
 fluctuations occurs only in the vicinity of $T_c$.

 The general features observed in the magnetic system are found also for the ferroelectric system. Starting from a macroscopically upolarised state at low temperature, the effect of increasing temperature is to increase the number of Type I vertices. As before, domains of Type I's form, separated by chains of Type II and III vertices. Effects of disorder are also completely analogous.

\begin{figure}
  \centering
  \includegraphics[width=0.5\columnwidth]{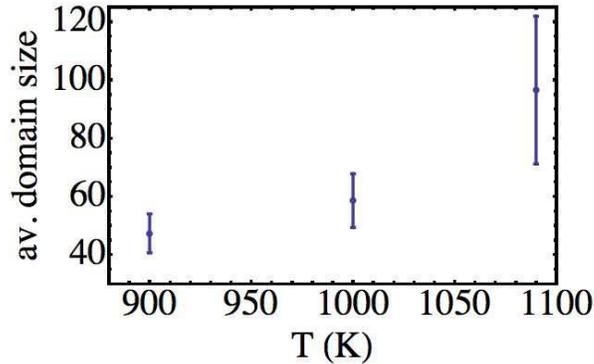}
   \caption{\label{dsize} The average size of a connected region (domain) of Type I vertices as a function of temperature as calculated using a self consistent Landau-Ginzburg model with parameters appropriate for PbTiO$_3$. The error bars show one standard deviation. }
  \end{figure}

 Finally, we note that domains grow at temperatures approaching $T_c$, as also found for the magnetic system. An example of average Type I domain size is shown in figure \ref{dsize} for three temperatures. The error bars correspond to the spread in sizes calculated over ten different initial configurations.

 %FIG example MF domain size

 It is useful to note that it is possible to define a relaxational dynamics based on the effective field $-\frac{\partial F}{\partial {\bf p}_i}$. The iteration process can thus be related directly to a real time dynamics. In this sense, the numerical iteration method is a time evolution, and these simulations give a feeling for the effect that different cooling rates may have on a system. If the system is cooled faster than the islands' polarisation can respond, then a higher number of
 unfavourable Type II and Type III vertices will be frozen into the artificial ferroelectric ice. This is interesting because these domain walls form without
 the presence of disorder and have a density depending on the history of the
system.

\end{document}